\documentclass[prd,aps,twocolumn,preprintnumbers,nofootinbib,showpacs,amsmath,amssymb,superscriptaddress]{revtex4-1}
\pdfoutput=1

\usepackage{color}
\definecolor{purple}{rgb}{0.7,0.0,0.5}
\usepackage[bookmarks = false, colorlinks = true, linkcolor = blue, citecolor = purple, urlcolor=purple]{hyperref}

\usepackage{amsfonts}
\usepackage{amsmath}
\usepackage{amssymb}
\usepackage{upgreek}
\usepackage{bm}
\usepackage{dcolumn}
\usepackage{epsfig}
\usepackage{graphicx}
\usepackage{graphics}
\usepackage[latin1]{inputenc}
\usepackage{latexsym}
\usepackage{rotating}
\usepackage{tipa}
\usepackage{hyperref}
\usepackage[all]{hypcap}
\usepackage{xspace}

\newcommand{\eg}{{\it e.g.~}}

\newcommand{\EE}{\mathcal{E}}

\newcommand{\TTT}{\mathcal{T}}

\newcommand{\NN}{\mathcal{N}}

\newcommand{\lsb}{\left[}
\newcommand{\rsb}{\right]}
\newcommand{\lb}{\left(}
\newcommand{\rb}{\right)}

\newcommand{\OO}{\mathcal{O}}
\newcommand{\mrm}{\mathrm}
\newcommand{\pd}{\partial}

\def\BS#1{{\textcolor{blue}{[Bogdan: #1]}}}

\newcommand {\bes} {\begin {equation*}}
\newcommand {\ees} {\end {equation*}}

\newcommand{\tr}{\mathop{\mathrm{tr}}}

\ifx\genfrac\sdflkaj

\else

\fi

\newcommand{\be}{\begin{eqnarray}}
\newcommand{\ee}{\end{eqnarray}}
\newcommand{\nn}{\nonumber}

\begin{document}

\title{Tomography from Entanglement}

\author{Jennifer Lin}
\affiliation{Enrico Fermi Institute and Department of Physics, University of Chicago, Chicago, IL 60637}

\author{Matilde Marcolli}
\affiliation{Department of Mathematics, California Institute of Technology, 253-37, Pasadena, CA 91125}

\author{Hirosi Ooguri}
\affiliation{Walter Burke Institute for Theoretical Physics, California Institute of Technology, 452-48, Pasadena, CA 91125}
\affiliation{Kavli Institute for the Physics and Mathematics of the Universe (WPI),
University of Tokyo, Kashiwa 277-8583, Japan}

\author{Bogdan Stoica}
\affiliation{Walter Burke Institute for Theoretical Physics, California Institute of Technology, 452-48, Pasadena, CA 91125}

\date{\today}

\preprint{CALT-TH 2014-162, IPMU14-0349}

\begin{abstract}

The Ryu-Takayanagi formula relates the entanglement entropy in a conformal field theory to the area of a minimal surface in its holographic dual. We show that this relation can be inverted for any state in the conformal field theory to compute the bulk stress-energy tensor near the boundary of the bulk spacetime, reconstructing the local data in the bulk from the entanglement on the boundary. We also show that positivity, monotonicity, and convexity of the relative entropy for small spherical domains between the reduced density matrices of any state and of the ground state of the conformal field theory, follow from positivity conditions on the bulk matter energy density. We discuss an information theoretical interpretation of the convexity in terms of the Fisher metric.

\end{abstract}

\pacs{11.25.Tq}

\maketitle

\section{Introduction}

Gauge/gravity duality posits an exact equivalence between certain conformal field theories (CFT's) with many degrees of freedom and higher dimensional theories with gravity. 
It is of obvious interest to understand how bulk spacetime geometry and gravitational dynamics emerge from a non-gravitating theory. 
In recent years, there have appeared hints that quantum entanglement plays a key role. One important development in this direction was the proposal of Ryu and Takayanagi \cite{Ryu:2006bv, Ryu:2006ef} that the entanglement entropy (EE) between a spatial domain $D$ of a CFT and its complement is equal to the area of the bulk extremal surface $\Sigma$ homologous to it, 
\be\label{rt}
S_{EE} = \min_{\partial D = \partial\Sigma}\frac{{\rm area}(\Sigma)}{4G_N}\,.
\ee
Using \eqref{rt}, \cite{Nozaki:2013vta, Blanco:2013joa, Lashkari:2013koa, Bhattacharya:2013bna, Faulkner:2013ica, Swingle:2014uza} showed the emergence of linearized gravity from entanglement physics of the CFT, as we review below.
In this note, we continue this program. We show that certain universal properties of CFT's, namely positivity, monotonicity and convexity of the relative entropy between the reduced density matrix on a small spherical domain of an excited state of the CFT and of its vacuum state, follow from positivity conditions on the matter energy density in the near-AdS region of its dual. Moreover, we show that the bulk stress-energy density in this region can be reconstructed point-by-point from the entanglement on the boundary.

We first set the stage with a review of concepts from quantum information theory before more precisely stating our main claim.

\subsection{Relative Entropy} 
 
{Relative entropy} (see \eg~\cite{Blanco:2013joa} and references therein) is a measure of distinguishability between two quantum states in the same Hilbert space. The relative entropy of two density matrices $\rho_0$ and $\rho_1$ is defined as
\be \label{rele}
S(\rho_1 | \rho_0) = \tr(\rho_1 \log \rho_1) - \tr(\rho_1 \log \rho_0)\,.
\ee
It is positive, and increases with system size:
\begin{eqnarray}
S(\rho_1|\rho_0) &\geq& 0\,, \label{positivity} \\
 S(\rho_1^W |\rho_0^W) &\geq&S(\rho_1^V|\rho_0^V) , \qquad W \supseteq V \,.
\label{mm1}
\end{eqnarray}
This second property is called monotonicity. 
When $\rho_0$ and $\rho_1$ are reduced density matrices on a spatial domain
$D$ for two states of a quantum field theory (QFT), which is the case we specialize to from this point on, \eqref{mm1} implies that $S(\rho_1|\rho_0)$ increases with the size of $D$. That is, over a family of scalable domains with characteristic size $R$, 
\be\label{monotonicity}
\partial_R S(\rho_1|\rho_0) \geq 0\,.
\ee

Defining the modular Hamiltonian $H_{mod}$ of $\rho_0$ implicitly through
\be\label{hmod}
\rho_0 = \frac{e^{-H_{mod}}}{\tr(e^{-H_{mod}})}\,,
\ee
it is easy to see that \eqref{positivity} is equivalent to 
\be\label{1stlaw}
S(\rho_1 |\rho_0) = \Delta \langle H_{mod}\rangle - \Delta S_{EE} \geq 0
\ee
where $\Delta\langle H_{mod}\rangle = \tr(\rho_1 H_{mod}) - \tr (\rho_0 H_{mod})$ is the change in the expectation value of the operator $H_{mod}$ \eqref{hmod} and $\Delta S_{EE} = -\tr(\rho_1 \log \rho_1) + \tr(\rho_0 \log \rho_0)$ is the change in the entanglement entropy across $D$ as one goes between the states.

When the states under comparison are close, the positivity \eqref{1stlaw} is saturated to leading order \cite{Blanco:2013joa,Lashkari:2013koa,Faulkner:2013ica}:
\be \label{1stlawi}
S(\rho_1 |\rho_0) = \Delta \langle H_{mod}\rangle - \Delta S_{EE}=0\,.
\ee
To see this, consider a reference state of the QFT characterized by $\rho_0$, and another, arbitrary state $\rho_1$. One can construct a family of interpolating density matrices 
\be \label{ldp}
\rho(\lambda) = \label{defl} (1-\lambda)\rho_0 + \lambda\rho_1\, ,
\ee
where $\lambda$ can be positive or negative.
Because the relative entropy $S(\rho_0|\rho(\lambda))$ is positive for either sign of $\lambda$,  the first derivative of this relative entropy with respect to $\lambda$ vanishes. This implies \eqref{1stlawi} to first order in $\lambda$.

Eq.~\eqref{1stlawi} is called the {entanglement first law} for its resemblance to the first law of thermodynamics. Indeed, when $\rho_0$ is a thermal density matrix $\rho_0 = e^{-\beta H}/\tr(e^{-\beta H})$, 
\eqref{1stlawi} reduces to $\Delta \langle H \rangle = T \Delta S$, an exact quantum version of the thermal first law. 

In general, the modular Hamiltonian \eqref{hmod} associated to a given density matrix is nonlocal. However, there are a few simple cases where it is explicitly known. When $\rho_0$ is the reduced density matrix of the vacuum state of a CFT on a disk of radius $R$ which (without loss of generality) we take to be centered at $\vec x_0 = 0$ \cite{Casini:2011kv},
\be \label{defhmod}
H_{mod} = \pi\intop_D d^{d-1}x\frac{R^2-\left|\vec x \right|^2}{R}T_{tt}(x)\,,
\ee
where $T_{tt}$ is the energy density of the CFT.

\subsection{Summary and Outline}

Our goal in this note is to use the entanglement in the CFT, in particular the relative entropy, to 
elucidate local physics in the bulk. Related recent works include 
an attempt to constrain the nonlinear gravitational equations of motion using the positivity of relative entropy \cite{Banerjee:2014oaa, Banerjee:2014ozp}, as well as the converse scenario of deriving differential equations constraining CFT entanglement from the nonlinear bulk Einstein equations
\cite{Bhattacharya:2012mia, Bhattacharya:2013bna, Nozaki:2013vta}.

Our starting point is a CFT whose vacuum state is dual to AdS$_{d+1}$.
We consider an arbitrary excited state of the CFT 
which has a semiclassical holographic bulk dual, whose metric can be parametrized as
\be\label{fge}
g_{AdS} = \frac{\ell_{AdS}^2}{z^2}\lsb dz^2 + \lb\eta_{\mu\nu}+ h_{\mu\nu}\rb dx^\mu dx^\nu\rsb\,.
\ee
Spacetime indices $a, b, \dots$ run over $(t, z, x^i)$ while $\mu, \nu, \dots$ run over $(t, x^i)$ and $i \in 1, \dots, d-1$ are boundary spatial directions. Moreover, we assume that the Ryu-Takayanagi formula holds in the excited state, and the relative entropy between the reduced density matrix $\rho_1$ of the excited state and $\rho_0$ of the ground state for the entangling disk $D$ of radius $R$ is computable using the formulae in the previous subsection. 

To apply a perturbative analysis in the bulk, we assume that the radius $R$ of the entangling domain is small compared to the typical energy scale $\EE \approx \langle T_{\mu\nu}\rangle^\frac{1}{d}$ of the state measured by the boundary stress tensor $T_{\mu\nu}$,
\be\label{condEV}
 \EE^d R^d \ll 1\,. 
\ee
Furthermore, we assume that the energy scales of the bulk fields are of the same order as the boundary stress-energy tensor energy scale, so that the geometry in the bulk region bounded by $D$ and the associated Ryu-Takayanagi surface is close to that of AdS.
%
%

In this limit, to order less than $\EE^{2d} R^{2d}$, we will show that the relative entropy is expressed as
\be\label{wec}
S(\rho_1|\rho_0) = 8\pi^2 G_N\intop_V \frac{R^2-(z^2+x^2)}{R} \varepsilon \sqrt{g_V} \, , 
\ee
where $G_N$ is Newton's constant,
 $V$ is a $d$-dimensional region on a constant-time slice bounded 
by the domain $D$ on the boundary  
and the Ryu-Takayanagi surface $\Sigma$ in the bulk, and $\sqrt{g_V}$ is the volume form in the bulk (including the time direction).
In particular, 
the positivity of the relative entropy follows the weak energy condition, i.e. the positivity of the bulk energy density $\varepsilon$ (see \eg \cite{hawking1973large}). Though the weak energy condition is not necessarily satisfied in AdS, it holds near the boundary of AdS, where we are evaluating (\ref{wec}).
We also note that the positivity is only required for the integrated quantity. 

In \cite{Faulkner:2013ica}, it was shown that  
the first law $S(\rho_1|\rho_0)=0$ in the linear approximation is equivalent 
to the linearized Einstein equation. This holds to the order $O(\EE^d R^d)$. 
Our result (\ref{wec}) improves
the approximation to the order less than $\EE^{2d} R^{2d}$ by taking into account the backreaction 
to the bulk stress tensor. 

Taking one derivative with respect to $R$, the monotonicity of the relative entropy can also be related to a bulk energy condition: 
\be 
\partial_R S(\rho_1 | \rho_0) = 8\pi^2 G_N\intop_V\left(1 + \frac{z^2+x^2}{R^2} \right)\varepsilon \sqrt{g_V}
 \,.
\ee
One more derivative relates the relative entropy to  
the integral of the energy density on $\Sigma$,
\be\label{mon}
\left(\partial^2_R + R^{-1} \partial_R - R^{-2} \right) S(\rho_1|\rho_0) = 16\pi^2 G_N \intop_\Sigma \varepsilon\sqrt{g_\Sigma} \,, \qquad
\ee
where $\sqrt{g_\Sigma}$ is the volume form on the Ryu-Takayanagi surface. We will show that 
\eqref{mon} can be inverted to express the bulk stress tensor point-by-point in the near-AdS region using the entanglement information of the CFT.

The outline of the paper is as follows. In Section \ref{bkgd} we review how to translate each quantity in \eqref{1stlaw} to holography, and sketch how to derive the linearized vacuum equations of motion from the entanglement first law using a generalized Stokes theorem argument \cite{Faulkner:2013ica, 1993Wald, Iyer:1994ys}. In Section \ref{main} we demonstrate that the positivity, monotonicity and convexity of the relative entropy for small spheres in the CFT are dual to the positivity conditions on the bulk stress tensor, and in Section \ref{ibi} we show how to invert (\ref{mon}) to obtain the bulk stress tensor locally in the near-AdS region. In Section \ref{ita}, we discuss to what extent we can recover from a general quantum-theoretic analysis the convexity of the relative entropy \eqref{mon} that was derived holographically. We comment on implications and open problems in Section \ref{disc}.

\section{Preliminaries}
\label{bkgd}

\subsection{$\Delta H_{mod}$ and $\Delta S_{EE}$ in holography}\label{dhmod}

We first review how each quantity appearing in the definition of the relative entropy \eqref{1stlaw} is mapped holographically. The modular Hamiltonian $H_{mod}$ for the reduced density matrix 
of the vacuum state of a CFT on the entangling disk $D$ of radius $R$, centered at a point on the boundary, is  a function of the CFT stress tensor $\langle T_{tt}\rangle $ \eqref{defhmod}. It vanishes in the CFT vacuum.  
In the excited state, $\langle T_{tt}\rangle$ in the CFT can be expressed as a function of bulk fields by using holographic renormalization (see \eg~\cite{Balasubramanian:1999re, deHaro:2000xn, Skenderis:2002wp}) or the shortcut of 
\cite{Faulkner:2013ica} to exploit the fact that the relative entropy in the CFT vanishes in the limit that the entangling domain shrinks to zero.
As long as the bulk matter fields contributing to $\langle T_{\mu\nu} \rangle$ are dual to operators 
with scaling dimension $\delta > d/2$, both methods give 
\be
\label{hmode}
\Delta \langle H_{mod}\rangle &=& \lim_{z \rightarrow 0} \frac{d\ell^2_{AdS}}{16 G_N} \intop_D d^{d-1}x \frac{R^2-|\vec x |^2}{R} z^{-d} \eta^{ij} h_{ij} \,. \qquad
\ee
In general, the right-hand side is modified by boundary counter terms if it involves operators with $\delta \leq d/2$. We hope to generalize our result to such a case in future work. 

The holographic EE in Einstein gravity is given by the Ryu-Takayanagi area formula \eqref{rt}.
On a constant time slice of pure AdS, the codimension-2 bulk extremal surface $\Sigma$
ending on a boundary sphere of radius $R$ is the half-sphere
\be\label{minads}
z_0(r) = \sqrt{R^2-r^2}\,.
\ee
The EE of the entangling disk of radius $R$ in the CFT vacuum is equal to the area functional of pure AdS evaluated on the surface \eqref{minads}.
Suppose we perturb the bulk metric away from pure AdS by $h_{ab}$ which is parametrically small. Because the original surface was extremal, the leading variation in the holographic EE comes from evaluating $z_0(r)$ \eqref{minads} on the perturbed area functional. One finds \cite{Lashkari:2013koa}
\be \label{seed}
\Delta S_{EE} = \frac{\ell^{d-1}_{AdS}
}{8 G_N R}\intop_{\Sigma} d^{d-1}x  (R^2 \eta^{ij} - x^ix^j )z^{-d}h_{ij}. \quad
\ee
At order $h^2$, one must account for corrections to the shape of the Ryu-Takayanagi
surface, see \eg\ \cite{Blanco:2013joa}.  

\subsection{Linearized Einstein Equations}\label{efe}

We now summarize the derivation of the linearized gravitational equations of motion from the entanglement first law \eqref{1stlawi}, as presented in \cite{Faulkner:2013ica}. 
%
The idea of \cite{Faulkner:2013ica} was to apply the Stokes theorem to the bulk $d$-dimensional region $V$ 
on a constant-time slice bounded by the entangling disk $D$ on the boundary and the extremal surface 
$\Sigma$ in the bulk. 
One can write $\Delta H_{mod}$ and $\Delta S_{EE}$ as integrals over $D$ and $\Sigma$ respectively of a local $d-1$ form $\pmb\chi$ that is a functional of the metric fluctuation $h_{ab}$. Within Einstein gravity, \cite{Faulkner:2013ica, 1993Wald, Iyer:1994ys} explicitly construct a $\pmb\chi[h_{ab}]$ that gives \eqref{hmode} and \eqref{seed} when integrated over $B$ and $\tilde B$, 
\be
\label{faulkner}
\intop_D \pmb\chi = \Delta\langle  H_{mod}\rangle , \quad \intop_{\Sigma}\pmb\chi = \Delta S_{EE}\,.
\ee 
Moreover, the exterior derivative of this $\pmb\chi$ is given by 
%
\be\label{dchi}
{\rm d}{\pmb\chi} = 2\xi^t  E^g_{tt}[h] g^{tt} \sqrt{g_V} dz\wedge dx^{i_1}\dots\wedge dx^{i_{d-1}},
\ee
where $\sqrt{g_V}$ is the natural volume form on $V$ induced from the bulk spacetime metric, and 
\be\label{kv}
\xi^a &=& -\frac{2\pi}{R}(t - t_0)[z\partial_z + x^i\partial_i] \\
&+& \frac{\pi}{R}[R^2-z^2-(t-t_0)^2- x^2]\partial_t \nn
\ee
is the Killing vector associated with $\Sigma$ \eqref{minads}, which is a bifurcate Killing horizon in pure AdS. $E^g_{ab}[h]$ are the linear gravitational equations of motion in vacuum.

By the Stokes theorem, 
the relative entropy is given by
\be\label{eqone}
S(\rho_1|\rho_0) = \Delta\langle H_{mod}\rangle - \Delta S_{EE} 
= \intop_V {\rm d}\pmb\chi\,. 
\ee
Considering \eqref{eqone} for every disk on a spatial slice at fixed time $t=0$, the entanglement first law $S(\rho_1|\rho_0)=0$
can then be shown to be equivalent to  $ E^g_{tt}[h] = 0$. 
Considering it for Lorentz-boosted frames gives vanishing of the other boundary components, $ E^g_{\mu\nu}[h] = 0$.
Finally, an argument appealing to the initial-value formulation gives vanishing of the remaining components of the linearized Einstein tensor that carry $z$ indices.

To summarize, \cite{Faulkner:2013ica} proved the existence of a $d-1$ form $\pmb\chi$ as a functional of a metric fluctuation $h_{ab}$, for which \eqref{faulkner} hold off shell and \eqref{dchi} holds with $E^g_{ab}[h]$ the linearized gravity equations of motion in vacuum.

By accounting for the $1/N$ correction to the Ryu-Takayanagi formula \cite{Faulkner:2013ana}, 
\cite{Swingle:2014uza} showed that the entanglement first law \eqref{1stlawi} implies 
the bulk linearized Einstein equations sourced by the {\it quantum} expectation value of 
the bulk stress-energy tensor, $\langle t_{ab}\rangle$. 
Assuming that the source of the linearized Einstein equation is a local QFT operator, one can then argue 
that the quantum expectation value appearing in their derivation can be uplifted to the bulk operator $t_{ab}$. 
In what follows, we remain in the large $N$ classical gravity limit, but assume 
the linearized Einstein equations sourced by the {\it classical} value of the bulk stress tensor and  use it to derive additional results.

\section{Effects due to Bulk Stress Tensor} \label{main}

We now evaluate the $(d-1)$-form $\pmb\chi$ of \cite{Faulkner:2013ica} on the bulk metric fluctuation $h_{ab}$ 
of the dual to an arbitrary excited state of a CFT, but in the interior of the Ryu-Takayanagi 
surface for the entangling disk \eqref{minads}, whose radius satisfies \eqref{condEV}. As the deviation of the bulk metric in the enclosed volume $V$ is parametrically small, all results of the above discussion carry over:
 \be\label{ineq}
\Delta S_{EE} - \Delta \langle H_{mod}\rangle
 = \intop_{\Sigma}\pmb\chi - \intop_D\pmb\chi = \intop_{\partial V}
\pmb\chi = \intop_V {\rm d}\pmb\chi\,. \quad
\ee
Here d$\pmb\chi$ is given by \eqref{dchi}, but $E^g_{ab}[h]$ is now the linearized vacuum Einstein tensor evaluated on  the $h_{ab}$ which is reconstructed from CFT data at non-linear level. 
This $E^g_{ab}[h_{ab}]$ is now not identically zero. 
Rather, the linearized Einstein tensor couples to bulk matter in the form of the bulk stress tensor
$t_{ab}$. Using
\be
E^g_{ab}[h_{ab}] = 8\pi G_N t_{ab}\,,
\ee
Eq. \eqref{ineq} becomes (see \eqref{dchi})
\be\label{ec}
S(\rho_1|\rho_0)  &=& \Delta H_{mod}-  \Delta S_{EE} \\
&=& 8\pi G \intop_V  \xi^t \epsilon \sqrt{g_V}\,,\nn
\ee
where the energy density $\epsilon$ appearing on the right corresponds to the $tt$-component of the stress-energy tensor, $\varepsilon = - t^{t}_{~t}$. 

For example, a massive scalar field in the bulk can contribute to the metric perturbation $h_{ab}$ as $\langle \OO \rangle^2 z^{2\Delta}$, where $\Delta$ is the scaling dimension of the corresponding operator on the boundary and $ \langle \OO \rangle $ is its expectation value, leading to an $O( \langle \OO \rangle^2 R^{2\Delta})$ effect in (\ref{ec}). On the other hand, corrections to the relative entropy by non-linear 
gravity effects are of the order $O(\EE^{2d}R^{2d})$ or higher, which we ignore. 
Thus, effects due to relevant operators with $\Delta < d$ are visible in our approximation. 

Now, we can show that the monotonicity \eqref{monotonicity} of the relative entropy
follows from a positivity condition on the bulk energy density. By taking a derivative of \eqref{ec} with respect to the radius $R$ of the entangling domain, we find
\be
\partial_R S(\rho_1|\rho_0) &=& 8\pi G_N \intop_{\Sigma}\xi^t \epsilon\sqrt{g_V}   \\
&+& 8\pi^2 G_N \intop_V \left(1 + \frac{x^2+z^2}{R^2} \right) \varepsilon\sqrt{g_V} \nn \\
\label{bulkintegral}
&=& 8\pi^2 G_N \intop_V \left(1 + \frac{x^2+z^2}{R^2} \right) \varepsilon\sqrt{g_V}.\ \ 
\ee
The integral over the Ryu-Takayanagi surface $\Sigma$ vanishes 
because $\xi^t$ \eqref{kv} vanishes on the surface. 
Assuming the weak energy condition ($i.e.$, the positivity of the energy density), we find the inequality
\be
\partial_R S(\rho_1|\rho_0)  \geq 0.
\ee
Though the weak energy condition is not necessarily satisfied in AdS, it holds near the boundary of AdS.
We also note that the positivity is only required for the integrated quantity. 

\section{Inverting the bulk integral}\label{ibi}

We found that $\partial_R S(\rho_1|\rho_0)$ is given by the integral of the energy density $\varepsilon$
over the region $V$ inside the Ryu-Takayanagi surface. We can invert this relation to compute 
$\varepsilon$ point-by-point in the bulk by using the relative entropy $S(\rho_1|\rho_0)$.

To show this, note that 
\be\label{rd}
\left( \partial_R + R^{-1} \right) S(\rho_1|\rho_0) = 16\pi^2 G_N\intop_V\varepsilon\sqrt{g_V}
\ee
so differentiating again,
\be\label{radon}
\left(\partial^2_R + R^{-1} \partial_R - R^{-2} \right) S(\rho_1|\rho_0) = 16\pi^2 G_N \intop_\Sigma \varepsilon\sqrt{g_\Sigma} \qquad
\ee
where $\sqrt{g_\Sigma}$ is the natural volume form on the Ryu-Takayanagi surface $\Sigma$
induced from the bulk
spacetime metric.
We note that the right-hand side is still non-negative if we assume the positivity of the 
bulk energy density. Thus, 
\be
\left(\partial^2_R + R^{-1} \partial_R - R^{-2} \right) S(\rho_1|\rho_0) \geq 0.
\ee

Here the bulk geometry is the unperturbed AdS, and its space-like section is the $d$-dimensional 
hyperbolic space. The surface $\Sigma$ is then totally geodesic.
In this case, the integral \eqref{radon} is the Radon transform and its inverse is known. For a smooth function $f$ on 
$d$-dimensional hyperbolic space, the Radon transform ${\cal R}f$ is an integral of $f$ over a $n$-dimensional 
geodesically complete submanifold with $n< d$. This gives a function on the space of geodesically complete 
submanifolds. The {\it dual} Radon transform ${\cal R}^* {\cal R}f$ gives back 
a function on the original hyperbolic space in the following way: pick a point in the hyperbolic space, consider
all geodesically complete submanifolds passing through the point, and integrate ${\cal R}f$ over such
submanifolds. 

It was shown by Helgason \cite{Helgason} that if $d$ is odd, $f$ is obtained by applying an appropriate differential operator on 
${\cal R}^* {\cal R}f$. We are interested in the case $n = d-1$ for which
\be
  f = \lsb (-4)^{(d-1)/2}\pi^{d/2-1}\Gamma(d/2) \rsb^{-1} Q({\bf \Delta}) {\cal R}^* {\cal R}f\,, \ 
\ee
where $Q({\bf \Delta}) $ is constructed from the Laplace-Beltrami operator ${\bf \Delta}$ on the hyperbolic space as
\be
Q({\bf \Delta}) &=& \lsb{\bf \Delta} + 1\cdot (d-2)\rsb\lsb{\bf \Delta}+2\cdot (d-3)\rsb \\
&\times& \cdots \times \lsb{\bf \Delta} + (d-2)\cdot 1\rsb \nn \, .
\ee

Applying this to \eqref{radon}, we find 
\be
\label{Helgason}
\varepsilon &=& \lsb (-4)^{(d+3)/2} \pi^{d/2+1}
\Gamma(d/2) G_N \rsb^{-1} \times \\ 
&\times& Q({\bf \Delta}) {\cal R}^* \left(\partial^2_R + 
R^{-1} \partial_R - R^{-2} \right) S(\rho_1|\rho_0) \,, \nn
\ee
when $d$ is odd. There is a similar formula when $d$ is even \cite{Rubin}.
The energy density is the time-time component of the stress-energy tensor $t_{tt}$. By computing the relative entropy in other Lorentz frames, we can also derive components $t_{\mu\nu}$ along the boundary. Finally, we can use the conservation law, $\nabla^a t_{ab}=0$, to obtain the remaining components, $t_{z\mu}, t_{\mu\nu}$. Thus,
we can use the entanglement data on the boundary to reconstruct all components of the bulk stress 
tensor. 

Since the Radon transform preserves positivity, the positivity of the energy density implies the positivity of $ \left(\partial^2_R + 
R^{-1} \partial_R - R^{-2} \right) S(\rho_1|\rho_0)$.
Conversely, the positivity of the latter implies the positivity of its dual Radon transform. 
It is interesting to note that $Q({\bf \Delta})$ in \eqref{Helgason} 
 is a positive definite operator when acting on normalizable functions on the hyperbolic space,
though this does not quite imply the positivity of the energy density.

\section{Comparison with Information Theoretic Analysis}\label{ita}
 
In this section, we discuss to what extent we can recover the monotonicity and convexity \eqref{mon} of the relative entropy from the following general property of the relative entropy. 
Consider a density matrix $\rho$ (with $\rho^*=\rho$, $\rho \geq 0$, and ${\rm tr}(\rho)=1$),
and two increments $h,\ell$, given by matrices with $h=h^*$, $\ell=\ell^*$ and ${\rm tr}(h)={\rm tr}(\ell)=0$. 
If the matrices $\rho,h,\ell$ satisfy $[\rho,h]=[\rho,\ell]=0$, then the relative entropy satisfies
\begin{equation}\label{Fisher}
S(\rho+h | \rho+\ell) \sim \langle (h-\ell), 
\frac{1}{2} \rho^{-1} (h-\ell)\rangle ,
\end{equation}
where the right-hand-side is the Fisher metric, with the
Hilbert--Schmidt inner product $\langle a, b \rangle ={\rm tr}(a^* b)$. Thus,
 the second order term is non-negative definite, and the quadratic form only vanishes for $h=\ell$.

The entanglement density matrices $\rho(R)$ and $\rho_0(R)$ discussed in this paper have
additional properties for small $R$. Since $H_{mod}$ is given by the integral (\ref{defhmod}) of $\xi^t T_{tt}$ over $|\vec x- \vec x_0|<R$, the Taylor expansion of $T_{tt}$ around $\vec x= \vec x_0$ gives 
$H_{mod} = h_0  R^d + \cdots$. Therefore, the density matrix
for the vacuum state can be expanded as
\be
\rho_0(R) = \frac{1}{\NN} - h'_0  R^d +\cdots \, ,
\ee
where  ${\rm tr}\ 1 = \NN$ and $h'_0= h_0 -\frac{1}{\NN} {\rm tr}\ h_0$ so that ${\rm tr}\ h'_0=0$. 
For $\rho(R)$, we postulate 
\be
\rho(R) = \frac{1}{\NN} + \sum_i \ell_i  R^{\delta_i} + h R^d + \cdots \, ,
\ee
so that
the small $R$ expansion of the relative entropy $S = \sum_i R^{2\delta_i} s_i + \cdots$ expected from the holographic computation above is reproduced. Here  ${\rm tr} \ \ell_i = 0$
and $\delta_i$'s are scaling dimensions of relevant operators, $\delta_i < d$.

The right-hand-side of (\ref{Fisher}) becomes
\be
\sum_{ij} \frac{\NN}{2}\, \langle \ell_i, \ell_j \rangle R^{\delta_i +
\delta_j} .
\ee
Thus, the leading order term of the relative entropy $S(\rho_1|\rho_0)$
can be estimated as
\be\label{info}
S(\rho_1|\rho_0) \sim \frac{\NN}{2} | \ell_1 |^2 R^{2\delta_1},
\ee
where $\delta_1=\min_i \{ \delta_i \}$.
Its first and second derivatives in $R$ have leading term
\be
 \partial_R S(\rho_1|\rho_0) &\sim& \NN \delta_1 | \ell_1 |^2
R^{2\delta_1 -1}  \,,\quad\ \\
 \left(\partial^2_R + R^{-1} \partial_R - R^{-2} \right)
S(\rho_1|\rho_0) &\sim&\frac{\NN}{2} |\ell_1 |^2 \times \\
&\times& (4 \delta_1^2-1) R^{2\delta_1-2} \, . \nn
\ee
The first is manifestly positive, and the second is non-negative provided 
$\delta_1 \geq 1/2$, which is satisfied by our assumption $\delta_1 > d/2$
for $d \geq 2$.\\

Our holographic analysis shows that the positivity and the convexity of the relative entropy hold
for subleading terms up to $O(R^{2d})$. On the other hand, corrections to (\ref{info}) may  
involve not only quadratic terms with $\delta_i + \delta_j < 2d$, but also cubic terms with 
$\delta_i + \delta_j + \delta_k < 2d$, $etc$. It appears that additional assumptions on the density matrices 
are required to explain the convexity from this point of view.

\section{Discussion} 
\label{disc}

We conclude with a few comments on prospects for future work.

In this paper, we have focused on bulk theories of classical Einstein gravity. However, it is probable that our result can be extended to higher-derivative classical gravities, as \cite{Faulkner:2013ica} has constructed the equivalent of the $d-1$ form ${\pmb\chi}$ in such theories. Another interesting question concerns generalizing away from the classical limit. If we were to add the $1/N$ correction to the Ryu-Takayanagi formula \cite{Faulkner:2013ana}, we would appear to obtain constraints on the quantum energy density from the positivity of the relative entropy. We leave these analyses to future work.

Another obvious question involves going beyond the small $R$ limit \eqref{condEV} that we have taken in this paper. 


\bigskip
\noindent
{\bf Acknowledgements}

We thank N. Bao, D. Harlow, T. Hartman, P. Hayden, N. Hunter-Jones, C. Keller, D. Kutasov, H. Liu, Y. Nakayama, 
S. Pufu, P. Sulkowski, T. Takayanagi, M. Van Raamsdonk and E. Witten for useful discussion. 
JL acknowledges support from the Sidney Bloomenthal fellowship at the University of Chicago. 
MM is currently supported by NSF grants PHY-1205440,
DMS-1201512, and DMS-1007207.
HO and BS are supported in part by the Walter Burke Institute for Theoretical Physics at Caltech, by U.S. DOE grant DE-SC0011632, and by a Simons Investigator award from the Simons Foundation. The work of HO is also supported in part by the WPI Initiative of MEXT of Japan, and JSPS Grant-in-Aid for Scientific Research C-26400240. He also thanks the hospitality of the Aspen Center for Physics and the National Science Foundation, which supports the Center under Grant No. PHY-1066293. BS is supported in part by a Dominic Orr Graduate Fellowship. JL, HO and BS would like to thank the Institute for Advanced Study, Princeton University, and the Simons Center for Geometry and Physics for hospitality. JL also thanks Caltech for hospitality.

\bibliographystyle{apsrev4-1}
\bibliography{HEE}

\begin{thebibliography}{21}%
\makeatletter
\providecommand \@ifxundefined [1]{%
 \@ifx{#1\undefined}
}%
\providecommand \@ifnum [1]{%
 \ifnum #1\expandafter \@firstoftwo
 \else \expandafter \@secondoftwo
 \fi
}%
\providecommand \@ifx [1]{%
 \ifx #1\expandafter \@firstoftwo
 \else \expandafter \@secondoftwo
 \fi
}%
\providecommand \natexlab [1]{#1}%
\providecommand \enquote  [1]{``#1''}%
\providecommand \bibnamefont  [1]{#1}%
\providecommand \bibfnamefont [1]{#1}%
\providecommand \citenamefont [1]{#1}%
\providecommand \href@noop [0]{\@secondoftwo}%
\providecommand \href [0]{\begingroup \@sanitize@url \@href}%
\providecommand \@href[1]{\@@startlink{#1}\@@href}%
\providecommand \@@href[1]{\endgroup#1\@@endlink}%
\providecommand \@sanitize@url [0]{\catcode `\\12\catcode `\$12\catcode
  `\&12\catcode `\#12\catcode `\^12\catcode `\_12\catcode `\%12\relax}%
\providecommand \@@startlink[1]{}%
\providecommand \@@endlink[0]{}%
\providecommand \url  [0]{\begingroup\@sanitize@url \@url }%
\providecommand \@url [1]{\endgroup\@href {#1}{\urlprefix }}%
\providecommand \urlprefix  [0]{URL }%
\providecommand \Eprint [0]{\href }%
\providecommand \doibase [0]{http://dx.doi.org/}%
\providecommand \selectlanguage [0]{\@gobble}%
\providecommand \bibinfo  [0]{\@secondoftwo}%
\providecommand \bibfield  [0]{\@secondoftwo}%
\providecommand \translation [1]{[#1]}%
\providecommand \BibitemOpen [0]{}%
\providecommand \bibitemStop [0]{}%
\providecommand \bibitemNoStop [0]{.\EOS\space}%
\providecommand \EOS [0]{\spacefactor3000\relax}%
\providecommand \BibitemShut  [1]{\csname bibitem#1\endcsname}%
\let\auto@bib@innerbib\@empty
\bibitem [{\citenamefont {Ryu}\ and\ \citenamefont
  {Takayanagi}(2006{\natexlab{a}})}]{Ryu:2006bv}%
  \BibitemOpen
  \bibfield  {author} {\bibinfo {author} {\bibfnamefont {S.}~\bibnamefont
  {Ryu}}\ and\ \bibinfo {author} {\bibfnamefont {T.}~\bibnamefont
  {Takayanagi}},\ }\href {\doibase 10.1103/PhysRevLett.96.181602} {\bibfield
  {journal} {\bibinfo  {journal} {Phys.Rev.Lett.}\ }\textbf {\bibinfo {volume}
  {96}},\ \bibinfo {pages} {181602} (\bibinfo {year} {2006}{\natexlab{a}})},\
  \Eprint {http://arxiv.org/abs/hep-th/0603001} {arXiv:hep-th/0603001 [hep-th]}
  \BibitemShut {NoStop}%
\bibitem [{\citenamefont {Ryu}\ and\ \citenamefont
  {Takayanagi}(2006{\natexlab{b}})}]{Ryu:2006ef}%
  \BibitemOpen
  \bibfield  {author} {\bibinfo {author} {\bibfnamefont {S.}~\bibnamefont
  {Ryu}}\ and\ \bibinfo {author} {\bibfnamefont {T.}~\bibnamefont
  {Takayanagi}},\ }\href {\doibase 10.1088/1126-6708/2006/08/045} {\bibfield
  {journal} {\bibinfo  {journal} {JHEP}\ }\textbf {\bibinfo {volume} {0608}},\
  \bibinfo {pages} {045} (\bibinfo {year} {2006}{\natexlab{b}})},\ \Eprint
  {http://arxiv.org/abs/hep-th/0605073} {arXiv:hep-th/0605073 [hep-th]}
  \BibitemShut {NoStop}%
\bibitem [{\citenamefont {Nozaki}\ \emph {et~al.}(2013)\citenamefont {Nozaki},
  \citenamefont {Numasawa}, \citenamefont {Prudenziati},\ and\ \citenamefont
  {Takayanagi}}]{Nozaki:2013vta}%
  \BibitemOpen
  \bibfield  {author} {\bibinfo {author} {\bibfnamefont {M.}~\bibnamefont
  {Nozaki}}, \bibinfo {author} {\bibfnamefont {T.}~\bibnamefont {Numasawa}},
  \bibinfo {author} {\bibfnamefont {A.}~\bibnamefont {Prudenziati}}, \ and\
  \bibinfo {author} {\bibfnamefont {T.}~\bibnamefont {Takayanagi}},\ }\href
  {\doibase 10.1103/PhysRevD.88.026012} {\bibfield  {journal} {\bibinfo
  {journal} {Phys.Rev.}\ }\textbf {\bibinfo {volume} {D88}},\ \bibinfo {pages}
  {026012} (\bibinfo {year} {2013})},\ \Eprint {http://arxiv.org/abs/1304.7100}
  {arXiv:1304.7100 [hep-th]} \BibitemShut {NoStop}%
\bibitem [{\citenamefont {Blanco}\ \emph {et~al.}(2013)\citenamefont {Blanco},
  \citenamefont {Casini}, \citenamefont {Hung},\ and\ \citenamefont
  {Myers}}]{Blanco:2013joa}%
  \BibitemOpen
  \bibfield  {author} {\bibinfo {author} {\bibfnamefont {D.~D.}\ \bibnamefont
  {Blanco}}, \bibinfo {author} {\bibfnamefont {H.}~\bibnamefont {Casini}},
  \bibinfo {author} {\bibfnamefont {L.-Y.}\ \bibnamefont {Hung}}, \ and\
  \bibinfo {author} {\bibfnamefont {R.~C.}\ \bibnamefont {Myers}},\ }\href
  {\doibase 10.1007/JHEP08(2013)060} {\bibfield  {journal} {\bibinfo  {journal}
  {JHEP}\ }\textbf {\bibinfo {volume} {1308}},\ \bibinfo {pages} {060}
  (\bibinfo {year} {2013})},\ \Eprint {http://arxiv.org/abs/1305.3182}
  {arXiv:1305.3182 [hep-th]} \BibitemShut {NoStop}%
\bibitem [{\citenamefont {Lashkari}\ \emph {et~al.}(2014)\citenamefont
  {Lashkari}, \citenamefont {McDermott},\ and\ \citenamefont
  {Van~Raamsdonk}}]{Lashkari:2013koa}%
  \BibitemOpen
  \bibfield  {author} {\bibinfo {author} {\bibfnamefont {N.}~\bibnamefont
  {Lashkari}}, \bibinfo {author} {\bibfnamefont {M.~B.}\ \bibnamefont
  {McDermott}}, \ and\ \bibinfo {author} {\bibfnamefont {M.}~\bibnamefont
  {Van~Raamsdonk}},\ }\href {\doibase 10.1007/JHEP04(2014)195} {\bibfield
  {journal} {\bibinfo  {journal} {JHEP}\ }\textbf {\bibinfo {volume} {1404}},\
  \bibinfo {pages} {195} (\bibinfo {year} {2014})},\ \Eprint
  {http://arxiv.org/abs/1308.3716} {arXiv:1308.3716 [hep-th]} \BibitemShut
  {NoStop}%
\bibitem [{\citenamefont {Bhattacharya}\ and\ \citenamefont
  {Takayanagi}(2013)}]{Bhattacharya:2013bna}%
  \BibitemOpen
  \bibfield  {author} {\bibinfo {author} {\bibfnamefont {J.}~\bibnamefont
  {Bhattacharya}}\ and\ \bibinfo {author} {\bibfnamefont {T.}~\bibnamefont
  {Takayanagi}},\ }\href {\doibase 10.1007/JHEP10(2013)219} {\bibfield
  {journal} {\bibinfo  {journal} {JHEP}\ }\textbf {\bibinfo {volume} {1310}},\
  \bibinfo {pages} {219} (\bibinfo {year} {2013})},\ \Eprint
  {http://arxiv.org/abs/1308.3792} {arXiv:1308.3792 [hep-th]} \BibitemShut
  {NoStop}%
\bibitem [{\citenamefont {Faulkner}\ \emph {et~al.}(2014)\citenamefont
  {Faulkner}, \citenamefont {Guica}, \citenamefont {Hartman}, \citenamefont
  {Myers},\ and\ \citenamefont {Van~Raamsdonk}}]{Faulkner:2013ica}%
  \BibitemOpen
  \bibfield  {author} {\bibinfo {author} {\bibfnamefont {T.}~\bibnamefont
  {Faulkner}}, \bibinfo {author} {\bibfnamefont {M.}~\bibnamefont {Guica}},
  \bibinfo {author} {\bibfnamefont {T.}~\bibnamefont {Hartman}}, \bibinfo
  {author} {\bibfnamefont {R.~C.}\ \bibnamefont {Myers}}, \ and\ \bibinfo
  {author} {\bibfnamefont {M.}~\bibnamefont {Van~Raamsdonk}},\ }\href {\doibase
  10.1007/JHEP03(2014)051} {\bibfield  {journal} {\bibinfo  {journal} {JHEP}\
  }\textbf {\bibinfo {volume} {1403}},\ \bibinfo {pages} {051} (\bibinfo {year}
  {2014})},\ \Eprint {http://arxiv.org/abs/1312.7856} {arXiv:1312.7856
  [hep-th]} \BibitemShut {NoStop}%
\bibitem [{\citenamefont {Swingle}\ and\ \citenamefont
  {Van~Raamsdonk}(2014)}]{Swingle:2014uza}%
  \BibitemOpen
  \bibfield  {author} {\bibinfo {author} {\bibfnamefont {B.}~\bibnamefont
  {Swingle}}\ and\ \bibinfo {author} {\bibfnamefont {M.}~\bibnamefont
  {Van~Raamsdonk}},\ }\href@noop {} {\  (\bibinfo {year} {2014})},\ \Eprint
  {http://arxiv.org/abs/1405.2933} {arXiv:1405.2933 [hep-th]} \BibitemShut
  {NoStop}%
\bibitem [{\citenamefont {Casini}\ \emph {et~al.}(2011)\citenamefont {Casini},
  \citenamefont {Huerta},\ and\ \citenamefont {Myers}}]{Casini:2011kv}%
  \BibitemOpen
  \bibfield  {author} {\bibinfo {author} {\bibfnamefont {H.}~\bibnamefont
  {Casini}}, \bibinfo {author} {\bibfnamefont {M.}~\bibnamefont {Huerta}}, \
  and\ \bibinfo {author} {\bibfnamefont {R.~C.}\ \bibnamefont {Myers}},\ }\href
  {\doibase 10.1007/JHEP05(2011)036} {\bibfield  {journal} {\bibinfo  {journal}
  {JHEP}\ }\textbf {\bibinfo {volume} {1105}},\ \bibinfo {pages} {036}
  (\bibinfo {year} {2011})},\ \Eprint {http://arxiv.org/abs/1102.0440}
  {arXiv:1102.0440 [hep-th]} \BibitemShut {NoStop}%
\bibitem [{\citenamefont {Banerjee}\ \emph
  {et~al.}(2014{\natexlab{a}})\citenamefont {Banerjee}, \citenamefont
  {Bhattacharyya}, \citenamefont {Kaviraj}, \citenamefont {Sen},\ and\
  \citenamefont {Sinha}}]{Banerjee:2014oaa}%
  \BibitemOpen
  \bibfield  {author} {\bibinfo {author} {\bibfnamefont {S.}~\bibnamefont
  {Banerjee}}, \bibinfo {author} {\bibfnamefont {A.}~\bibnamefont
  {Bhattacharyya}}, \bibinfo {author} {\bibfnamefont {A.}~\bibnamefont
  {Kaviraj}}, \bibinfo {author} {\bibfnamefont {K.}~\bibnamefont {Sen}}, \ and\
  \bibinfo {author} {\bibfnamefont {A.}~\bibnamefont {Sinha}},\ }\href
  {\doibase 10.1007/JHEP05(2014)029} {\bibfield  {journal} {\bibinfo  {journal}
  {JHEP}\ }\textbf {\bibinfo {volume} {1405}},\ \bibinfo {pages} {029}
  (\bibinfo {year} {2014}{\natexlab{a}})},\ \Eprint
  {http://arxiv.org/abs/1401.5089} {arXiv:1401.5089 [hep-th]} \BibitemShut
  {NoStop}%
\bibitem [{\citenamefont {Banerjee}\ \emph
  {et~al.}(2014{\natexlab{b}})\citenamefont {Banerjee}, \citenamefont
  {Kaviraj},\ and\ \citenamefont {Sinha}}]{Banerjee:2014ozp}%
  \BibitemOpen
  \bibfield  {author} {\bibinfo {author} {\bibfnamefont {S.}~\bibnamefont
  {Banerjee}}, \bibinfo {author} {\bibfnamefont {A.}~\bibnamefont {Kaviraj}}, \
  and\ \bibinfo {author} {\bibfnamefont {A.}~\bibnamefont {Sinha}},\
  }\href@noop {} {\  (\bibinfo {year} {2014}{\natexlab{b}})},\ \Eprint
  {http://arxiv.org/abs/1405.3743} {arXiv:1405.3743 [hep-th]} \BibitemShut
  {NoStop}%
\bibitem [{\citenamefont {Bhattacharya}\ \emph {et~al.}(2013)\citenamefont
  {Bhattacharya}, \citenamefont {Nozaki}, \citenamefont {Takayanagi},\ and\
  \citenamefont {Ugajin}}]{Bhattacharya:2012mia}%
  \BibitemOpen
  \bibfield  {author} {\bibinfo {author} {\bibfnamefont {J.}~\bibnamefont
  {Bhattacharya}}, \bibinfo {author} {\bibfnamefont {M.}~\bibnamefont
  {Nozaki}}, \bibinfo {author} {\bibfnamefont {T.}~\bibnamefont {Takayanagi}},
  \ and\ \bibinfo {author} {\bibfnamefont {T.}~\bibnamefont {Ugajin}},\ }\href
  {\doibase 10.1103/PhysRevLett.110.091602} {\bibfield  {journal} {\bibinfo
  {journal} {Phys.Rev.Lett.}\ }\textbf {\bibinfo {volume} {110}},\ \bibinfo
  {pages} {091602} (\bibinfo {year} {2013})},\ \Eprint
  {http://arxiv.org/abs/1212.1164} {arXiv:1212.1164} \BibitemShut {NoStop}%
\bibitem [{\citenamefont {Hawking}\ and\ \citenamefont
  {Ellis}(1973)}]{hawking1973large}%
  \BibitemOpen
  \bibfield  {author} {\bibinfo {author} {\bibfnamefont {S.}~\bibnamefont
  {Hawking}}\ and\ \bibinfo {author} {\bibfnamefont {G.}~\bibnamefont
  {Ellis}},\ }\href {http://books.google.com/books?id=QagG\_KI7Ll8C} {\emph
  {\bibinfo {title} {The Large Scale Structure of Space-Time}}},\ Cambridge
  Monographs on Mathematical Physics\ (\bibinfo  {publisher} {Cambridge
  University Press},\ \bibinfo {year} {1973})\BibitemShut {NoStop}%
\bibitem [{\citenamefont {{Wald}}(1993)}]{1993Wald}%
  \BibitemOpen
  \bibfield  {author} {\bibinfo {author} {\bibfnamefont {R.~M.}\ \bibnamefont
  {{Wald}}},\ }\href {\doibase 10.1103/PhysRevD.48.R3427} {\bibfield  {journal}
  {\bibinfo  {journal} {\prd}\ }\textbf {\bibinfo {volume} {48}},\ \bibinfo
  {pages} {3427} (\bibinfo {year} {1993})},\ \Eprint
  {http://arxiv.org/abs/gr-qc/9307038} {gr-qc/9307038} \BibitemShut {NoStop}%
\bibitem [{\citenamefont {Iyer}\ and\ \citenamefont
  {Wald}(1994)}]{Iyer:1994ys}%
  \BibitemOpen
  \bibfield  {author} {\bibinfo {author} {\bibfnamefont {V.}~\bibnamefont
  {Iyer}}\ and\ \bibinfo {author} {\bibfnamefont {R.~M.}\ \bibnamefont
  {Wald}},\ }\href {\doibase 10.1103/PhysRevD.50.846} {\bibfield  {journal}
  {\bibinfo  {journal} {Phys.Rev.}\ }\textbf {\bibinfo {volume} {D50}},\
  \bibinfo {pages} {846} (\bibinfo {year} {1994})},\ \Eprint
  {http://arxiv.org/abs/gr-qc/9403028} {arXiv:gr-qc/9403028 [gr-qc]}
  \BibitemShut {NoStop}%
\bibitem [{\citenamefont {Balasubramanian}\ and\ \citenamefont
  {Kraus}(1999)}]{Balasubramanian:1999re}%
  \BibitemOpen
  \bibfield  {author} {\bibinfo {author} {\bibfnamefont {V.}~\bibnamefont
  {Balasubramanian}}\ and\ \bibinfo {author} {\bibfnamefont {P.}~\bibnamefont
  {Kraus}},\ }\href {\doibase 10.1007/s002200050764} {\bibfield  {journal}
  {\bibinfo  {journal} {Commun.Math.Phys.}\ }\textbf {\bibinfo {volume}
  {208}},\ \bibinfo {pages} {413} (\bibinfo {year} {1999})},\ \Eprint
  {http://arxiv.org/abs/hep-th/9902121} {arXiv:hep-th/9902121 [hep-th]}
  \BibitemShut {NoStop}%
\bibitem [{\citenamefont {de~Haro}\ \emph {et~al.}(2001)\citenamefont
  {de~Haro}, \citenamefont {Solodukhin},\ and\ \citenamefont
  {Skenderis}}]{deHaro:2000xn}%
  \BibitemOpen
  \bibfield  {author} {\bibinfo {author} {\bibfnamefont {S.}~\bibnamefont
  {de~Haro}}, \bibinfo {author} {\bibfnamefont {S.~N.}\ \bibnamefont
  {Solodukhin}}, \ and\ \bibinfo {author} {\bibfnamefont {K.}~\bibnamefont
  {Skenderis}},\ }\href@noop {} {\bibfield  {journal} {\bibinfo  {journal}
  {Commun.Math.Phys.}\ }\textbf {\bibinfo {volume} {217}},\ \bibinfo {pages}
  {595} (\bibinfo {year} {2001})}\BibitemShut {NoStop}%
\bibitem [{\citenamefont {Skenderis}(2002)}]{Skenderis:2002wp}%
  \BibitemOpen
  \bibfield  {author} {\bibinfo {author} {\bibfnamefont {K.}~\bibnamefont
  {Skenderis}},\ }\href {\doibase 10.1088/0264-9381/19/22/306} {\bibfield
  {journal} {\bibinfo  {journal} {Class.Quant.Grav.}\ }\textbf {\bibinfo
  {volume} {19}},\ \bibinfo {pages} {5849} (\bibinfo {year} {2002})},\ \Eprint
  {http://arxiv.org/abs/hep-th/0209067} {arXiv:hep-th/0209067 [hep-th]}
  \BibitemShut {NoStop}%
\bibitem [{\citenamefont {Faulkner}\ \emph {et~al.}(2013)\citenamefont
  {Faulkner}, \citenamefont {Lewkowycz},\ and\ \citenamefont
  {Maldacena}}]{Faulkner:2013ana}%
  \BibitemOpen
  \bibfield  {author} {\bibinfo {author} {\bibfnamefont {T.}~\bibnamefont
  {Faulkner}}, \bibinfo {author} {\bibfnamefont {A.}~\bibnamefont {Lewkowycz}},
  \ and\ \bibinfo {author} {\bibfnamefont {J.}~\bibnamefont {Maldacena}},\
  }\href {\doibase 10.1007/JHEP11(2013)074} {\bibfield  {journal} {\bibinfo
  {journal} {JHEP}\ }\textbf {\bibinfo {volume} {1311}},\ \bibinfo {pages}
  {074} (\bibinfo {year} {2013})},\ \Eprint {http://arxiv.org/abs/1307.2892}
  {arXiv:1307.2892} \BibitemShut {NoStop}%
\bibitem [{\citenamefont {Helgason}(1959)}]{Helgason}%
  \BibitemOpen
  \bibfield  {author} {\bibinfo {author} {\bibfnamefont {S.}~\bibnamefont
  {Helgason}},\ }\href@noop {} {\bibfield  {journal} {\bibinfo  {journal} {Acta
  Math.}\ }\textbf {\bibinfo {volume} {102}},\ \bibinfo {pages} {239} (\bibinfo
  {year} {1959})}\BibitemShut {NoStop}%
\bibitem [{\citenamefont {Rubin}(2002)}]{Rubin}%
  \BibitemOpen
  \bibfield  {author} {\bibinfo {author} {\bibfnamefont {B.}~\bibnamefont
  {Rubin}},\ }\href@noop {} {\bibfield  {journal} {\bibinfo  {journal} {Adv.
  Math}\ }\textbf {\bibinfo {volume} {170}},\ \bibinfo {pages} {206} (\bibinfo
  {year} {2002})}\BibitemShut {NoStop}%
\end{thebibliography}%

\end{document}